\documentclass[twocolumn,showpacs,amsmath,amssymb,aps,prd]{revtex4}

\usepackage{bm}

\begin{document}

\title{Rotational terms and quantum degeneracy in black holes}

\author{Kostiantyn  Ropotenko}
 \email{ro@stc.gov.ua, ropotenko@i.ua}

\affiliation{State Administration of Communications, Ministry of
Transport and Communications of Ukraine,\\ 22, Khreschatyk, 01001,
Kyiv, Ukraine}

\date{\today}

\begin{abstract}
It is believed that the first law of black-hole mechanics has no
independent physical significance and acquires it only after
identifying with the first law of thermodynamics. It is argued here
that the first law of black-hole mechanics has a direct physical
significance: not only the term $\Omega dJ$ but all its terms have
the same mechanical meaning - the rotational kinetic energy of a
black hole in real or in an internal space. Moreover, it is shown
that the Kerr-Newman black hole is a system of non-degenerate plane
rotators represented by the corresponding terms in the first law of
black-hole mechanics. It is found that a degeneracy arises because
the energy of a black hole does not depend on where an internal
angular momentum of a black hole associated with the black hole area
is determined on the horizon.

\end{abstract}

\maketitle

\section{Introduction}

The statistical interpretation of the black-hole entropy of
Bekenstein and Hawking
\begin{equation}
\label{in1} S_{\rm BH}=\frac{A}{4 l_P^{2}}
\end{equation}
remains a central problem in black hole physics \cite{all}. The
majority of approaches to the problem can be reduced to the
following main principles suggested by Bekenstein as early as $38$
years ago \cite{bek}:
\begin{itemize}
\item The black hole is a composite object.
\item Constituents of a black hole responsible for black hole entropy
reside on its event horizon; the area of the horizon is quantized
\begin{equation}
\label{in2} A_n=\Delta A \cdot n,\quad n=0,1,2,...,
\end{equation}
where $\Delta A $ is the quantum of black hole area, $\Delta A \sim
l_P^{2}$, so that the horizon surface consists of $n$ identical
patches - constituents of a black hole - each of area $\Delta A $.
Then, if every patch can have $k$ states, the total number of states
of a black hole is
\begin{equation}
\label{in3} W=k^n.
\end{equation}
\item From the statistical mechanics point of view, the black hole is a
conventional object; in particular, the entropy of a black hole is
the logarithm of the number of states associated with its horizon,
\begin{equation}
\label{in4} S = \ln W = n\ln k.
\end{equation}
Because the patches are independent of one another, the total
entropy of a black hole is just $n$ times the entropy of a single
patch $s_1=\ln k$.
\item The area spectrum (\ref{in2}) implies a discrete mass spectrum
with the level spacing
\begin{equation}
\label{in5} \Delta M_n \sim \frac{1}{M_n}.
\end{equation}
\end{itemize}
For convenience in the subsequent comparison, we shall refer to the
approach (model) based on these principles as \emph{the composite
black hole approach} (\emph{model}). However these principles raise
new puzzles. A puzzle is \cite{sred1}, that locally there is nothing
special about the horizon, so it is hard to see why it should behave
as an object with the local degrees of freedom (\ref{in3}). Another
puzzle is that we assign an entropy to something which is a
classical solution of the gravitational field equations, something
which behaves like a soliton, but we do not usually assign an
entropy to a soliton \cite{hoo}, \cite{sus1}. One more puzzle is
that the black hole, although additive (\ref{in4}), is nevertheless
an indivisible object. The next puzzle is related with the spectrum
(\ref{in5}). It implies a discrete emission spectrum. But if the
black hole is a conventional object, then the separation between
subsequent energy levels should be exponentially small $\Delta M_n
\sim \exp (-S_{\rm BH})$ and the emission spectrum is practically
continuous. Finally, the greatest puzzle is the nature of black hole
constituents and their quantum states. In connection with these
$38$-year puzzles a question arises: Is it possible that this is not
the point that there is not yet quantum theory of gravity, but the
point is that our model is not completely adequate to the physical
nature of black holes?

In \cite{ro1} a new approach to the problem of black hole entropy
was developed and applied to a Schwarzschild black hole. It is based
on the concept of internal angular momentum of a black hole
$L_z=A/(8\pi G)$. This approach does not need the concept of
constituents for determining the black hole entropy and is free of
the difficulties of the composite black hole approach. The basic
idea of the approach will be presented below. For convenience, we
shall refer to this approach (model) as \emph{the fundamental black
hole approach} (\emph{model}). The purpose of this paper is to
summarize and extend this approach to a generic Kerr-Newman black
hole. The Kerr-Newman case is more difficult than the Schwarzschild
one. The point is that a Kerr-Newman black hole has additional
degrees of freedom related with rotation and electromagnetism. If
the black hole is a fundamental object, then these degrees of
freedom can give a contribution to the total degeneracy of the black
hole. Note that the statistical interpretation of a Kerr-Newman
black hole has been studied by many authors but in the framework of
the composite black hole model (the list of references is too long
to be presented here). Moreover, these authors ignored the problem
of rotation of a symmetric structureless black hole in real space
from the quantum mechanical point of view.

The main results of the paper are as follows.
\begin{itemize}
\item It is believed that the first law of black-hole mechanics
\begin{equation}
\label{in6} dM=\frac{k}{8\pi G}dA+\Omega dJ+\Phi dQ,
\end{equation}
has no independent physical significance and acquires it only after
identifying with the first law of thermodynamics. It is argued that
the first law of black-hole mechanics has a direct physical
significance: not only the term $\Omega dJ$ but \emph{all} terms in
the law have the same mechanical meaning - the rotational kinetic
energy of a black hole in real or internal space.
\item A black hole has rotational symmetries and no internal structure.
Classically it can rotate about any of its axes (the Kerr solution).
But from a quantum mechanical point of view such a motion is
impossible. It is argued that the rotation of a black hole is
nevertheless observable due to the dragging effect. Moreover, it is
the dragging effect, that removes the degeneracy with respect to the
direction of $J$.
\item At first sight it seems that in the fundamental black hole approach
all three terms in (\ref{in6}) should give a contribution in the
total degeneracy factor of a black hole. But this is not so. The
Kerr-Newman black hole is a system of plane rotators represented by
the corresponding terms in the first law of black-hole mechanics.
That is, the first law of black-hole mechanics should read as
\begin{equation}
\label{in7} dM=\omega_1 dL_{1z}+\omega_2 dL_{2z}+\omega_3 dL_{3z},
\end{equation}
rather than (\ref{in6}). Here $\omega_i$ are the angular frequencies
and $L_{iz}$ are the $z$-components of the angular momenta of the
black hole in the corresponding spaces. The energy levels of
rotators are \emph{non-degenerate}.
\item The \emph{degeneracy} of energy levels of a black hole
arises because the energy of a black hole does not depend on where
$L_{1z}$ is determined on the horizon. Quantization of
$L_{1z}=A/(8\pi G)$ gives the equidistant area spectrum of a black
hole with the area quantum $\Delta A = 8\pi l_P^{2}$. Since the
precision with which $L_{1z}$ can be determined equals to the size
of the area quantum $\Delta A = 8\pi l_P^{2}$, the degeneracy of a
black hole is
\begin{equation}
\label{in8} W = \frac{A}{8\pi l_P^{2}}.
\end{equation}
That is why it is determined by the first term in black hole
mechanics.
\item The horizon is not a configuration surface with local degrees of freedom but
\emph{phase space} (\emph{surface}). Once we have accepted this, the
black hole entropy is no longer arbitrary and uniquely determined by
\begin{equation}
\label{in9} S =2\pi W.
\end{equation}
\end{itemize}

The organization of the paper is as follows. In Sec. II we consider
the degeneracy in Kerr-Newman black holes. We begin with the basic
idea suggested early in \cite{ro1}. Then we consider a trouble with
the first law of black-hole mechanics and argue that the law has
independent physical significance. We also show that the Kerr-Newman
black hole can be viewed as a system of non-degenerate plane
rotators. Finally, we find the origin of black hole degeneracy. In
Sec. III we consider some applications of our approach.

\section{The black hole degeneracy}
\subsection{The basic idea}

In \cite{ro2} an internal angular momentum of a Schwarzschild black
hole $L_z=A/(8\pi G)$ was determined. Quantization of $L_z$ gives
the equidistant area spectrum of a black hole $A_m=\Delta A \cdot
m,\quad m=0,1,2,...,$ with the area quantum $\Delta A = 8\pi
l_P^{2}$. The number of microstates is intrinsically an integer. But
$\exp (2\pi m)$ is not integral \cite{mag}. On the other hand, the
energy of a black hole does not depend on where $L_z$ is located on
the horizon. The precision with which $L_z$ can be determined equals
the size of the area quantum. Therefore, the number of states
accessible to a black hole can be determined as $m = A/8\pi
l_P^{2}$. However, once we have accepted this, the black hole
entropy is uniquely determined by $S=2\pi \mbox{(number of states)}$
(not by $S=\log \mbox{(number of states)}$). This means that the
black hole is a nonadditive object. This agrees with the
thermodynamical properties of a black hole. In particular, the
Bekenstein-Hawking entropy is not a homogeneous first order function
of the black hole energy. Moreover, we cannot divide a black hole
into two independent subsystems by a partition as an ideal gas in a
box (the area theorem). And the black hole constituents cannot be
extracted from a black hole. Therefore the black hole cannot be
thought as made up of any constituent subsystems each of them
endowed with its own independent thermodynamics \cite{arc}. We have
to consider a single black hole as a whole system. On the other
hand, the black hole is a vacuum solution of the gravitational
equations and can be viewed as a kind of gravitational soliton i.e.
as a physical object, localized "within the event horizon" and
possessing a mass $M$.  Moreover, for $r> R_g$ the Schwarzschild
coordinate $t$ is timelike and the coordinate $r$ is spacelike. But
for $r < R_g$, $t$ is a spacelike coordinate and $r$ is timelike. So
"time" and "radial" coordinates swap character when we cross $r =
R_g$. This looks like a twist of spacetime. Thus the  black hole
should be viewed as a fundamental object like an elementary particle
or, rather, a string (because we can assign an entropy to the
string). Of course, this idea is not new but rather folklore. 't
Hooft \cite{hoo} had already pointed out that there is no
fundamental difference between black holes and elementary particles,
and Susskind and other researchers deepened this insight still
further \cite{sus1}. According to this point of view the spectrum of
particles does not terminate at the Planck mass but continues on to
indefinitely large mass in the form of black holes. It is necessary
to stress, however, that in our approach we do not postulate the
fundamental character of black holes  but rather derive it.

\subsection{The degeneracy problem}

The Kerr-Newman case is more difficult than the Schwarzschild one.
The fact is that a Kerr-Newman black hole has additional degrees of
freedom associated with rotation and electromagnetism. The first law
of black hole thermodynamics and the generalized Smarr formula are
(neglecting a hypothetical magnetic charge)
\begin{equation}
\label{deg1} dM=T dS+\Omega dJ+\Phi dQ,
\end{equation}
\begin{equation}
\label{deg2} M=2TS+2\Omega J+\Phi Q,
\end{equation}
where $M$ is the black-hole mass, $T$ is the Hawking temperature,
$S$ is the Bekenstein-Hawking entropy, $\Omega$ is the angular
velocity, $J$ is the angular momentum, $\Phi$ is the electric
potential and $Q$ is the electric charge of a black hole. According
to statistical physics \cite{lan1}, the entropy of a conventional
composite body is a function of its internal energy alone. If the
black hole is such an object, then its entropy is
\begin{equation}
\label{deg3} S=S(M-2\Omega J-\Phi Q),
\end{equation}
and the additional degrees of freedom do not contribute to the black
hole degeneracy. In statistical physics the internal energy of a
body is interpreted as the total of the kinetic and potential energy
of all the constituents that compose it. If the black hole is a
fundamental object, then this concept has no meaning. In this case
$M$ is an eigenvalue of a single-particle Hamiltonian and the
entropy is a function of the total energy
\begin{equation}
\label{deg4} S=S(M).
\end{equation}
The degeneracy of the energy level $M$ would then be a product of
three degeneracy factors, one depending only on $A$, one only on $J$
(usually $2J+1$), and third only on $Q$. Is this true? To answer the
question, we should investigate the nature of terms $k (dA/8\pi G)$,
$\Omega dJ$ and $\Phi dQ$ in the first law of black hole mechanics.

\subsection{The trouble with the first law of black-hole mechanics}

It is widely believed that the laws of black hole mechanics have no
independent physical significance and acquire it only after
identifying with the laws of thermodynamics. For example, the first
law of black-hole mechanics reads
\begin{equation}
\label{law1} dM=\frac{k}{8\pi G}dA+\Omega dJ+\Phi dQ.
\end{equation}
It is nothing but a statement of energy conservation for a black
hole. Here, $dM$, is the change in mass (energy) of a black hole.
The second and third terms on the right hand side are usually
interpreted as changes in the energy due to rotation and
electromagnetism. But the first term with $dA$ does not have a
direct physical interpretation. As is well known, all basic
conservation laws of energy also contain a mix of terms of different
nature, but all these terms have clear physical meaning. In contrast
to this, the first law of black-hole mechanics has no independent
physical significance and becomes meaningful after identifying with
the first law of thermodynamics if one assumes the following
expressions for the temperature and entropy of a black hole
\begin{equation}
\label{law2} T_{\rm H}=\frac{k}{2\pi}, \quad S_{\rm BH}=\frac{A}{4
l_P^{2}}.
\end{equation}
The same is true, of course, of the generalized Smarr formula
(\ref{deg2}). In what follows it will be convenient to use the
formula
\begin{equation}
\label{law3} M=2k\frac{A}{8\pi G}+2\Omega J+2 \Theta Q^{2},
\end{equation}
where
\begin{equation}
\label{law4} \Theta=\frac{\Phi}{2Q},
\end{equation}
and the surface gravity $k$, the angular velocity $\Omega$, and the
electric potential $\Phi$ are
\begin{equation}
\label{law5} k=\frac{r_+-M}{r_+^{2}+a^{2}},\quad
\Omega=\frac{a}{r_+^{2}+a^{2}},\quad
\Phi=\frac{Qr_+}{r_+^{2}+a^{2}},
\end{equation}
Here $r_+=M+\sqrt{M^{2}-Q^{2}-a^{2}}$ and $a$ is the specific
angular momentum, $a=J/M$. As is easily seen, all three terms in
(\ref{law3}) have the same structure of the form $2\omega_i L_i$,
where $\omega_1 =k$, $\omega_2 =\Omega$, $\omega_3 =\Theta$ and $L_1
=A/(8\pi G)$, $L_2 =J$, $L_3 =Q^{2}$. The factors $\omega_i$ are in
turn very similar and have the same dimensions of frequency. The
$L_i$ have dimensions of action. What does this mean?

\subsection{An assumption}

It turns out that the first law of black-hole mechanics has a direct
physical significance: not only the term $\Omega dJ$ but \emph{all}
terms in the first law of black-hole mechanics have the same
mechanical meaning - the rotational energy of a black hole in real
or in an internal space. Namely, we assume that the Kerr-Newman
black hole can be viewed as a system of plane rotators represented
by the corresponding terms in the first law of black-hole mechanics.
That is, the first law of black-hole mechanics should read as
\begin{equation}
\label{as1} dM=\omega_1 dL_{1z}+\omega_2 dL_{2z}+\omega_3
dL_{3z}=\sum_i \omega_i dL_{iz}.
\end{equation}
rather than (\ref{law1}). Here $\omega_i$ are the angular
frequencies and $L_{iz}$ are the $z$-components of angular momenta
of a black hole in real or in an internal space. Their concrete
expressions will be determined below. Accordingly
\begin{equation}
\label{as2} M=2\omega_1 L_{1z}+2\omega_2 L_{2z}+2\omega_3
L_{3z}=2\sum_i \omega_i L_{iz}.
\end{equation}
Moreover, the energy levels of these rotators are non-degenerate, so
that the degeneracy factors associated with $L_{1z}$, $L_{2z}$ and
$L_{3z}$ do not give a contribution to the total degeneracy of a
black hole. There are good reasons for supposing that this is in
fact so.

\subsection{Rotational terms: $2 k (A/8\pi G)$}

The black hole entropy is a concept defined in the rest frame of an
external fiducial observer. Moreover, the Kerr-Newman metric
extended into the region within the event horizon cannot describe
the spacetime inside the black hole \cite{lan2}. So, since we do not
deal with the interior of a black hole, we regard the Euclidean
formulation as the more fundamental one. After all calculations with
the Euclideanized Kerr-Newman have been completed, we analytically
continue the results obtained back to real value of coordinate time
$t$. Moreover, since the black hole entropy is associated with the
event horizon, we deal only with the Rindler section of the whole
Euclidean Kerr-Newman manifold. The important fact is that in the
near-horizon approximation the metric of an arbitrary black hole can
be reduced to the Rindler form.

For example, the Rindler section of the Euclidean Schwarzschild
manifold $R^{2}\times S^{2}$ is an analytic continuation of that
part of the Lorentzian geometry that just lies outside or at the
event horizon, $r\geq 2GM$. In transforming from Schwarzschild to
Euclidean Rindler coordinates the Schwarzschild time $t$ transforms
to a variable $\varphi =kt$ in the Euclidean Rindler plane $R^{2}$,
$k=1/(4GM)$. But the metric has a coordinate singularity
corresponding to $r = 2GM$. Regularity is obtained if $\varphi$ is
interpreted as an angular coordinate with periodicity $2\pi$. In
\cite{ro2}, it was shown that there exists the $z$ component of the
angular momentum which is conjugate to this angle,
$\hat{L}_z=i\hbar\partial/\partial \varphi$ with an eigenvalue
$L_z=A/8\pi G$. We call it Rindler angular momentum. In the
Euclidean formulation the Rindler angular momentum and the Hawking
temperature have the same origin - the periodicity in the
Schwarzschild imaginary time - and, therefore, the universal
geometrical nature. The angular momentum $\hat{L}_z$ is the
generator of rotations around $z$ axis (this axis corresponds to
$r=2GM$). Since there exists only one way of rotation, the black
hole represents \emph{a plane rotator} in an internal space. As
early as $38$ years ago, by proving that the black hole horizon area
is an adiabatic invariant, Bekenstein showed \cite{bek} that the
quantum of black hole area is of the form $\Delta A = 8\pi l_P^{2}$.
In \cite{ro2}, by following the approach used by Susskind
\cite{sus1} to derive the Rindler energy, quantization of the black
hole area (and entropy) was obtained from the commutation relation
and quantization condition for $L_z$. Namely, since
\begin{equation}
\label{ar1} L_z=\frac{A}{8\pi G}=m\hbar,\quad m=0,1,2,...\,,
\end{equation}
then
\begin{equation}
\label{ar2} A=8\pi l_P^{2}\cdot m,
\end{equation}
and
\begin{equation}
\label{ar3} S=2\pi \cdot m.
\end{equation}
The energy of a Schwarzschild black hole in terms of a plane rotator
is
\begin{equation}
\label{ar4} M=2\omega L_z=2\hbar \omega \cdot m,
\end{equation}
where $\omega=k$ is the frequency of a black hole in an internal
space associated with $L_z$. Note that $L_z$ is an adiabatic
invariant as well as $A$.

In \cite{ro1} it was suggested that quantization of a Schwarzschild
black hole is nothing but the Landau quantization and a
Schwarzschild black hole represents a two-dimensional isotropic
oscillator with an additional interaction $\omega \hat{L}_z$, where
$\omega=k$. But the model of an oscillator presupposes a zero-point
energy of a black hole in the absence of the black hole which seems
unlikely. In contrast, the spectrum of a plane rotator does not
contain a zero-point energy. In what follows we shall use the model
of a plane rotator. The model of a harmonic oscillator can be
considered only as an approximation; the model reduces to that of a
plane rotator in the limit of large $m$.

Setting $t=-i\tau$ and reasoning as in the the Schwarzschild case,
we can define the Euclidean Kerr-Newman metric \cite{haw1}. This
metric is complex and it is asymptotically flat in a coordinate
system rotating with the angular velocity of a black hole $\Omega$.
The locus $r_+=M+\sqrt{M^{2}-Q^{2}-a^{2}}$ will be a conical
singularity unless we identify the point $(\tau, r, \theta, \phi)$
with the point $(\tau+i2\pi k^{-1}, r, \theta, \phi+i2\pi \Omega
k^{-1})$. As a result, the real coordinate time $t$ transforms to an
angular coordinate about the "axis"
$r_+=M+\sqrt{M^{2}-Q^{2}-a^{2}}$. Alternatively we can in addition
analytically continue in the specific angular momentum $a$ and
charge $Q$ to get a real Riemannian metric. In this case the
Euclidean Kerr-Newman manifold has a structure identical to that of
the Euclidean Schwarzschild manifold $R^{2}\times S^{2}$ and, in
particular, it covers only the exterior of a black hole and it also
requires the compactification of the Euclidean time in order to
eliminate a conical singularity. In an exactly similar manner to the
Schwarzschild case, we can determine Rindler angular momentum of a
Kerr-Newman black hole with the quantized eigenvalues \cite{ro2}
\begin{equation}
\label{ar5} L_{1z}\equiv\frac{A}{8\pi G}=m\hbar,\quad m=0,1,2,...\,,
\end{equation}
where $A$ is the area of a Kerr-Newman black hole. Since the
Euclidean Kerr-Newman spacetime has no region corresponding to the
region $r<r_+$ in the Lorentzian spacetime, the negative integers
$m$ are ruled out (as mentioned in the beginning, continuation of
the Lorentzian Kerr-Newman line element inside the surface of the
horizon has no physical meaning at all). As in the Schwarzschild
case, we can consider the Kerr-Newman black hole as \emph{a plane
rotator} associated with the angular momentum. Analogously, the
first term on the right hand side in (\ref{law3}) can be regarded as
the rotational energy of a Kerr-Newman black hole
\begin{equation}
\label{ar6} 2k\frac{A}{8\pi G} = 2\omega_1 L_{1z},
\end{equation}
where $\omega_1= k$ is the angular frequency of the black hole in an
internal space. As is well known, the energy levels of a plane
rotator except the ground state is doubly degenerate (this follows
from the fact that for every energy level the rotator can rotate
both in positive and in the negative direction). Since there are
only positive values of $m$, the plane rotator does not give a
contribution to the total degeneracy of a black hole.

\subsection{Rotational terms: $2 \Omega J$}

At first glance the rotation of a black hole in real space is
impossible. A Schwarzschild black hole is spherical in shape apart
from quantum fluctuations and has no internal structure.
Classically, it can rotate about any of its axes. Analogously, a
classical Kerr-Newman black hole can rotate around its axis of
symmetry. But from a quantum mechanical point of view the rotation
of such objects is unobserved. A black hole cannot rotate, because
any rotation leaves its horizon surface invariant and thus by
definition does not change the quantum-mechanical state; there is
nothing "inside a black hole" to change its position during the
rotation and there is nothing marked on its surface to define the
orientation. Moreover, no orbiting spots can observed on the
horizon, since all radiation from the horizon is infinitely
redshifted. But our arguments "against the rotation" are imperfect.
Indeed, in Newton's theory a gravitational field is in no way
dependent on the motion of matter. So the gravitational fields of a
rotating sphere and the same sphere at rest are completely
identical. But in Einstein's theory this is not so: a rotating
sphere drags spacetime around itself. This phenomenon is known as
the dragging of inertial frames. It is also known as the
Lense-Thirring effect. The effect is differential, stronger near the
black hole and weaker at larger distances. Moreover, within the
ergosphere the dragging is so strong that no object can remain at
rest. It is the dragging effect, that makes the rotation of a black
hole observable. Thus, illuminating space by a suitable beam of test
particles we may detect rotation of a black hole. And the degree to
which the particles rotate is a measure of how rapidly this hole
rotates.

The Kerr-Newman black hole is axially symmetric but not spherically
symmetric (i.e. rotationally symmetric about one axis only which is
the angular-momentum axis). Since the angular velocity $\Omega$ is
constant over the horizon, the black hole rotates rigidly. Therefore
we can regard the Kerr-Newman black hole as a rigid symmetric
rotator or top. According to quantum mechanics \cite{lan3}, the
stationary states of a symmetrical top are described by three
quantum numbers: the angular momentum $J$ ($J=0,1,2,...$) and its
components along the axis of the top $J_\zeta = K$
($K=J,J-1,...,-J$) and along the $z$-axis fixed in space $J_z$
($J_z=J,J-1,...,-J$). The degeneracy of each energy level of the top
with $K\neq 0$ is $W=2(2J+1)$ because $J_z$ can take $2J+1$
different values for a given value of $J$, and $K$ can be either
positive or negative, corresponding to the two possible directions
of rotation about the axis of angular momentum.

But we have ignored the frame-dragging effect. It turns out that
when it is taken into account, the degeneracy factor of a
Kerr-Newman black hole is completely removed. To deduce this, it is
not in fact necessary to solve the equation $\hat{J}\psi=J\psi$ in
the Kerr-Newman metric; instead, we can argue as follows. First, a
black hole drags a fixed coordinate system about its axis of angular
momentum. As a result, the stationary states of our symmetrical top
are determined only by the quantum number $J_z$. Secondly, the
hole's rotation drags the coordinate system into orbital motion in
the \emph{same} direction as the hole rotates. As a result, the
two-fold degeneracy with respect to values of $J_z$ are completely
removed. As is well known, in order to compensate the dragging
effect and to have a convenient family of observers for which events
at the same (Boyer-Lindquist) time are simultaneous, a family of
zero angular momentum observers is used. These observers are at
fixed $r$ and $\theta$, but have a constant angular velocity
$\omega=d\phi/dt$. An observer corotating with the frame-dragging
angular velocity $\omega$ is in a state of zero angular momentum,
and experiences no centrifugal forces. From the point of view of
such an observer a Kerr-Newman black hole is viewed as a
non-degenerate plane rotator rather than a symmetrical top. Thus the
second term on the right hand side in (\ref{law3}) is the kinetic
energy of rotation of a Kerr-Newman black hole in real space,
\begin{equation}
\label{rot1} 2\Omega dJ = 2\omega_2 L_{2z},
\end{equation}
where
\begin{equation}
\label{rot2} \omega_2= \Omega, \quad  L_{2z}= J_z.
\end{equation}

\subsection{Rotational terms: $2\Theta Q^{2}$}

The third term $2\Theta Q^{2}$ is the change in the electrostatic
energy of a black hole. This term is also related with rotations but
in an internal space. The electric charge is a conserved quantity.
This is a consequence of invariance of the Lagrangian under the
one-dimensional group $U(1)$ of gauge transformations. This group is
equivalent to $O(2)$, the orthogonal group of rotations in a plane.
Two-dimensional rotations or gauge transformations belong to the
Abelian group. As a result, the electric charge is additive in the
same way as the $z$ component of the angular momentum. On the other
hand, the electric charge is quantized. But it is not known for
certain why it is quantized. There have been many suggestions,
including Kaluza-Klein models \cite{kal}, magnetic monopoles
\cite{mon} and grand unified theories \cite{GUT} to explain the
quantization of electric charge. The important fact is that all
these suggestions are closely related  to the quantization of
angular momentum. For example, in the original Kaluza-Klein model
charged particles are ones that go round in the fifth curled up
dimension (neutral particles do not move in fifth dimension). The
charge is proportional to the angular momentum of the motion round
the curled up fifth dimension. In quantum theory, angular momentum
is quantized, so charge is quantized. Note that when a Kerr-Newman
black hole is described in the framework of a five-dimensional
Kaluza-Klein model, the quantities $\Omega$ and $\Phi$ enter the
expressions in a similar manner, and their properties are to a
certain extent similar \cite{bl}.

We shall assume that the electric charge of a black hole is an
integer multiple of the fundamental unit of electric charge $e$.
Since the charge operator $\hat{Q}$ is the generator of the $U(1)$
gauge transformations, its spectrum of eigenvalues should be of the
form
\begin{equation}
\label{cha1} Q= e n, \quad n=0,\pm1,\pm2,...
\end{equation}
Then
\begin{equation}
\label{cha2} 2\Theta Q^{2}=2 \Theta (4\pi\alpha n^{2}\hbar),
\end{equation}
where the definition of the fine structure constant $\alpha=
e^{2}/(4\pi \hbar)$ has been used. At this point we should note the
following. First, as has been stated above, there is an analogy
between the electric charge generating gauge transformations and the
$z$ component of the angular momentum generating rotations in a
plane. Secondly, as is easily seen from (\ref{cha2}), the states of
electrostatic energy of a black hole are doubly degenerate due to
the factor $n^{2}$. This is just the degeneracy we should expect for
a plane rotator. Therefore we can interpret the term $2\Theta Q^{2}$
as the kinetic rotational energy of a black hole in an internal
charge space and express it in terms of a plane rotator
\begin{equation}
\label{cha3} 2\Theta Q^{2}= 2 \omega_3 L_{3z}.
\end{equation}
Note that the interpretation of $2\Theta Q^{2}$ as the rotational
energy of a plane rotator has more formal significance than that of
$2k (A/8\pi G)$ and $2\Omega J$. Because of this, we do not
determine the frequency $\omega_3$ and the angular momentum
$L_{3z}$. Nevertheless, we shall use the form $2 \omega_3 L_{3z}$ to
uniform the term $2\Theta Q^{2}$ with other rotational terms.
Finally we must return to the degeneracy factor. The point is that
despite the existence of two kind of charges with opposite signs in
nature, the sign of the black hole charge $Q$ is given, so the
energy levels of the rotator are in fact non-degenerate.

\subsection{The origin of degeneracy}

As has been shown above, a Kerr-Newman black hole has three
rotational degrees of freedom associated with the angular momenta
$L_{1z}$, $L_{2z}$ and $L_{3z}$. But these degrees of freedom  do
not give a contribution to the total degeneracy of a black hole. We
did not determine $L_{3z}$ and $\omega_3$ explicitly. If we
determined $L_{3z}$ and $\omega_3$ as $L_{3z}=Q^{2}$ and $\omega_3=
\Theta$, for example, we should have an additional degeneracy
associated with $\omega_i$. It is clear, however, that in this case
all $\omega_i$ are linearly independent. So, is there a degeneracy
in black holes at all? It turns out that the energy of a Kerr-Newman
black hole, as in the Schwarzschild case, does not depend on where
$L_{1z}$ is located on the horizon. The precision with which
$L_{1z}$ can be determined equals the size of the area quantum
$\Delta A = 8\pi l_P^{2}$. Therefore, the number of states
accessible to a black hole can be determined as
\begin{equation}
\label{or1} W = \frac{A}{8\pi l_P^{2}}.
\end{equation}
We can imagine this as follows. Associate our rotator with a vortex
with the area of core $\Delta A = 8\pi l_P^{2}$. Then the number of
ways to place the vortex on the horizon is just (\ref{or1}), the
ratio of the area of a black hole to the area of the core.  As a
result, the (configuration) surface of the event horizon becomes
\emph{phase space} of a black hole, as it should; as is well known,
the phase space  of a plane rotator is two-dimensional. Therefore,
quantization of the black hole area and entropy is nothing but
quantization of the phase surface
\begin{equation}
\label{or2} S=2\pi m.
\end{equation}
Note that string theory needs to introduce the notion of a stretched
horizon to avoid the problem of the local degrees of freedom. In
loop quantum gravity, it is believed that the only possible degrees
of freedom on the horizon have to be global or topological,
described by a topological quantum field theory. According to the
traditional approach we would have to take the logarithm of $W$. But
in this case the generalized second law of black hole thermodynamics
(GSL) would be violated \cite{ro1} (it appears that much earlier,
Gour and Mayo showed \cite{gour} that the formula for the black hole
entropy $S=f(A)$ with the function $f(A)=\ln A$ clashes with the GSL
and must be excluded). It turns out that once we have accepted the
fact that the black hole degeneracy is proportional to the area, the
black hole entropy is no longer arbitrary and uniquely determined by
\begin{equation}
\label{or3} S=2\pi W
\end{equation}
(not by $S=\log W$). This relation is the only way to reconcile the
formula $S=2\pi m$ with the requirement that $W$ be integral. The
absence of the logarithm means that the black hole is a nonadditive
object. Reasoning as in the the Schwarzschild case (subsection A),
we conclude that the Kerr-Newman black hole is a fundamental object.

\section{Some applications}

\subsection{System of black holes (and matter)}

To avoid misunderstanding, we should add a clarifying remark
concerning a system of black holes. Suppose, for simplicity, that
two black holes are far apart and their interaction is negligible,
so that they can be viewed as statistically independent. Let
$S_{1(2)}=2\pi m_{1(2)}$ and $W_{1(2)}=m_{1(2)}$ be the entropy and
degeneracy of the first (second) black hole, respectively. Then the
number of states for the combined system is $W=W_1W_2=m_1m_2$. What
is the entropy of the system? Obviously, we cannot write the total
entropy as $S=2\pi m_1 m_2$ because our system is not a single black
hole. It seems that we would take the logarithm of $W$: $\ln W = \ln
m_1 + \ln m_2$. But in this case, as mentioned above, we cannot
interpret $\ln m_{1(2)}$ as the entropy of the first (second) black
hole. Does $W=m_1m_2$ exist? Yes, it does. But $\ln m_{1(2)}$ is not
the entropy of the first (second) black hole. Despite this failure
the laws of thermodynamics are still valid, so we can define the
entropy as
\begin{equation}
\label{sys1} S=S_1+S_2=2\pi (m_1+m_2).
\end{equation}
At this point we refer back to the meaning of the quantum number
$m$. It is the angular ("magnetic") quantum number. Since
interaction is weak, the angular momentum $L_z$ of the whole system
can be regarded as the sum of the angular momenta $L_{1z}$ and
$L_{2z}$ of its parts,
\begin{equation}
\label{sys2} L_z=L_{1z}+L_{2z}.
\end{equation}
Then it follows that
\begin{equation}
\label{sys3} m=m_1+m_2.
\end{equation}
According to thermodynamics the entropy of combine system is
additive. Thus the law of addition for the Rindler angular momenta
is nothing but a law of thermodynamics. We have considered the case
of two independent black holes. But we can extend it to an arbitrary
number of black holes (and matter).

Note that the relation $S=2\pi m$ can be also viewed as the angular
momentum quantization condition on the phase of wave-function: if
the eigenfunction of $L_z$ is to be single-valued, it must be
periodic in phase, with period $2\pi$. So we can consider $m$ also
as a topological number (winding number). As has been shown above,
it is an additive number.

\subsection{The mean separation between energy levels}

From (\ref{ar4}) it follows that the separation between energy
levels of a black hole is
\begin{equation}
\label{sep1} \Delta M_n=\frac{m_P^{2}}{2M_n}.
\end{equation}
This value is equal in order of magnitude to the width of energy
level $ R_g^{-1}$. This value, however, does not agree with
estimation obtained from the usual definition of entropy, $\langle
\Delta M_n\rangle \sim \exp (-S_{\rm BH})$. As mentioned in
Introduction, this compounds a problem in the composite black hole
approach. The fact is that in this approach the energy levels should
split due to unavoidable interactions between the constituents, so
the mean separation between energy levels should be really $\sim
\exp (-S_{\rm bh})$. This implies that the discreteness of the
spectrum is very blurred and difficult to see observationally. In
contrast, in the fundamental black hole approach there are no
constituents, so there is no splitting and the discreteness can be
observed.

\subsection{The relation $L_z=\alpha'M^{2}$}

Although the black hole is a fundamental object, it is unstable and
decays by emitting Hawking radiation (as is well known, stability
does not appear to be a criterion of the fundamental nature).
Therefore black holes can be viewed as resonant poles in the
S-matrix for scattering of stable particles \cite{sred}. In this
case there there should be poles in the complex $s$ plane (where the
Mandelstam variable $s$ is the total center-of-mass energy squared)
at
\begin{equation}
\label{res1} s_m=M_m^{2}-i\Gamma_m,
\end{equation}
where $\Gamma_m$ is the width of energy level in the $s$ plane and
the energy spacing between the subsequent energy levels of a black
hole is given by (\ref{sep1}). Srednicki noted \cite{sred} that
(\ref{sep1}) is very unusual behavior for a set of resonances.
Namely, its series never terminates. As is well known, all known
laboratory systems (for example, such as nuclei) have dense
resonances, but always there is a threshold above which the poles
are replaced by a cut. It turns out that the strange behavior of
black hole resonances noted by Srednicki can be explained in our
model. The Schwarzschild black hole is "the ground state of the
Kerr-Newman black hole". So we restrict our consideration to
uncharged, non-rotating black holes. The point is that the Rindler
angular momentum $L_z= 2GM^{2}$ is proportional to the square of the
mass of a black hole and increases without limit
\begin{equation}
\label{res2} L_z= \alpha'M^{2}.
\end{equation}
Here we have introduced the notation $\alpha'=2m_P^{-2}$; the reason
for this will be clear in a moment. This resembles the well-known
angular momentum - mass relation for hadronic resonances. As is well
known, the graph of the angular momentum $J$ of hadronic resonances
against their mass squared falls into lines $J=\alpha' M^{2}$ called
Regge trajectories. The constant $\alpha'$ is known as the Regge
slope. Instead of terminating abruptly as in the case of nuclei, the
graph continue on indefinitely, implying that quarks don't fly apart
when spun too fast. In contrast to nuclei, there is no a threshold
there. This is a manifestation of quark confinement. It has a simple
explanation in the string picture where the relation $J=\alpha'
M^{2}$ emerges naturally from a rotating open string. In string
model, hadrons are modeled by relativistically rotating strings
capped with massless quarks at both ends. For example, meson
resonances obey the relation $J=\alpha' M^{2}$ with the slope
$\alpha'\sim (1 \mbox{ GeV})^{2}$. Here the slope $\alpha'$ is
inversely proportional to the string tension. But this is an
effective string theory. As is well known, to date string theory is
considered as the unified theory of particle physics and gravity, so
the slope of fundamental strings is determined by the fundamental
constants of gravity and quantum theory, $\alpha' \sim m_P^{-2}$.
But it is just the slope of a black hole (\ref{res2}).


\begin{thebibliography}{99}
\bibitem{mag}
There is an assumption that this is not always the case, because the
value $W=\exp(2\pi m)$ is approximate. An argument is (see, for
example, M. Maggiore, Phys. Rev. Lett. {\textbf{100}}, 141301
(2008)) that since the number of states is the exponential of a very
large number, we cannot calculate $W$ with a precision of order one,
necessary to distinguish an integer from a non-integer value. An
objection is that the entropy $S$ is a continuously variable whereas
$W$, being an integer, varies discontinuously. Therefore $W$ is to
be considered as just enormously large, so large that a change of
one unit in $W$ influences $S$ to only a vanishingly small degree.
In our opinion, this is not a problem of the precision of
calculation: we do not need, for example, to calculate $W=\exp(\ln
(2011))$ with a precision of order one, to know that $W$ is
integral.
\end{thebibliography}
\end{document}